\documentclass[]{spie}  
\usepackage[]{graphicx}
\usepackage[]{color}

\newcommand{\mum}{\mathrm{\mu m}}

\title{Performance of Hybrid NbTiN-Al Microwave Kinetic Inductance Detectors as Direct Detectors for Sub-millimeter Astronomy} 

\author{R.M.J.~Janssen\supit{a}, J.J.A.~Baselmans\supit{b}, A.~Endo\supit{a},  L.~Ferrari\supit{c}, S.J.C.~Yates\supit{c}, A.M.~Baryshev\supit{c,d}, T.M.~Klapwijk\supit{a,e}
\skiplinehalf
\supit{a}Kavli Institute of NanoScience, Faculty of Applied Sciences, Delft University of Technology, Lorentzweg 1, 2628 CJ Delft, The Netherlands\\
\supit{b}SRON, Sorbonnelaan 2, 3584 CA Utrecht, The Netherlands\\
\supit{c}SRON, Landleven 12, 9747 AD Groningen, The Netherlands\\
\supit{d}Kapteyn Astronomical Institute, University of Groningen, P.O. Box 800, 9700AV Groningen, The Netherlands\\
\supit{e}Physics Department, Moscow State Pedagogical University, Moscow 119991, Russia
}

\authorinfo{R.J.: E-mail: r.m.j.janssen@tudelft.nl, Telephone: $+$31 15 27 88106\\
Copyright 2014 Society of Photo-Optical Instrumentation Engineers. One print or electronic copy may be made for personal use only. Systematic reproduction and distribution, duplication of any material in this paper for a fee or for commercial purposes, or modification of the content of the paper are prohibited.}

\begin{document} 
  \maketitle 

\begin{abstract}
In the next decades millimeter and sub-mm astronomy requires large format imaging arrays and broad-band spectrometers to complement the high spatial and spectral resolution of the Atacama Large Millimeter/sub-millimeter Array. The desired sensors for these instruments should have a background limited sensitivity and a high optical efficiency and enable arrays thousands of pixels in size. Hybrid microwave kinetic inductance detectors consisting of NbTiN and Al have shown to satisfy these requirements. We present the second generation hybrid NbTiN-Al MKIDs, which are photon noise limited in both phase and amplitude readout for loading levels $P_{\mathrm{850GHz}} \geq 10$ fW. Thanks to the increased responsivity, the photon noise level achieved in phase allows us to simultaneously read out approximately 8000 pixels using state-of-the-art electronics. In addition, the choice of superconducting materials and the use of a Si lens in combination with a planar antenna gives these resonators the flexibility to operate within the frequency range $0.09 < \nu < 1.1$ THz. Given these specifications, hybrid NbTiN-Al MKIDs will enable astronomically usable kilopixel arrays for sub-mm imaging and moderate resolution spectroscopy.
\end{abstract}



\section{Introduction}
In the next decades ALMA\cite{Brown2004} will make a major contribution in the study of the millimeter and sub-mm universe thanks to its high sensitivity, spectral resolution and spatial resolution. However, ALMA has a maximum instantaneous bandwidth of $8$ GHz and a small $\sim 18''$ field of view. This limits its performance as a survey instrument. To complement ALMA large format imaging arrays are thus required\cite{Scott2010}. These instruments require tens of thousands of pixels to fill the complete field of view of existing and future single dish sub-mm telescope, such as APEX\cite{Gusten2006}, IRAM\cite{Baars1987} and CCAT\cite{Woody2012}. For minimum observation time each pixel should have background limited performance. In addition, it's greatly beneficial if the same technology can be employed within all the atmospheric windows in the sub-mm regime $(100 \ \mathrm{GHz} < \nu < 1 \ \mathrm{THz})$.\\
The most promising candidate to fulfil these requirements are microwave kinetic inductance detectors (MKIDs)\cite{Day2003} due to their inherent potential for frequency domain multiplexing. MKIDs are superconducting resonators, thousands of which can be coupled to a single feedline by varying their resonance frequency. Each resonator is sensitive to changes in the Cooper pair density induced by absorption of sub-mm radiation. By monitoring the change in either phase or amplitude of the complex feedline transmission at the MKID resonance one can measure the absorbed photon power.\\
Using antenna coupled MKIDs made from NbTiN and Al photon noise limited performance has been achieved down to 100 fW of loading by 350 GHz radiation in both readout modes.\cite{Janssen2013}. The design of the hybrid NbTiN-Al MKIDs aims to simultaneously maximize the phase response and minimize the two-level system (TLS) noise contribution\cite{Gao2007}. In these proceedings we present the second generation antenna-coupled hybrid NbTiN-Al MKIDs designed for ground-based sub-mm astronomy. We show that these devices achieve photon noise limited performance in both amplitude and phase readout when exposed to as little as 10 fW of 850 GHz radiation.
\section{Hybrid NbTiN-Al MKID Design}
   \begin{figure}[t]
   \begin{center}
   \begin{tabular}{c c c}
   \includegraphics[width = 0.45\textwidth]{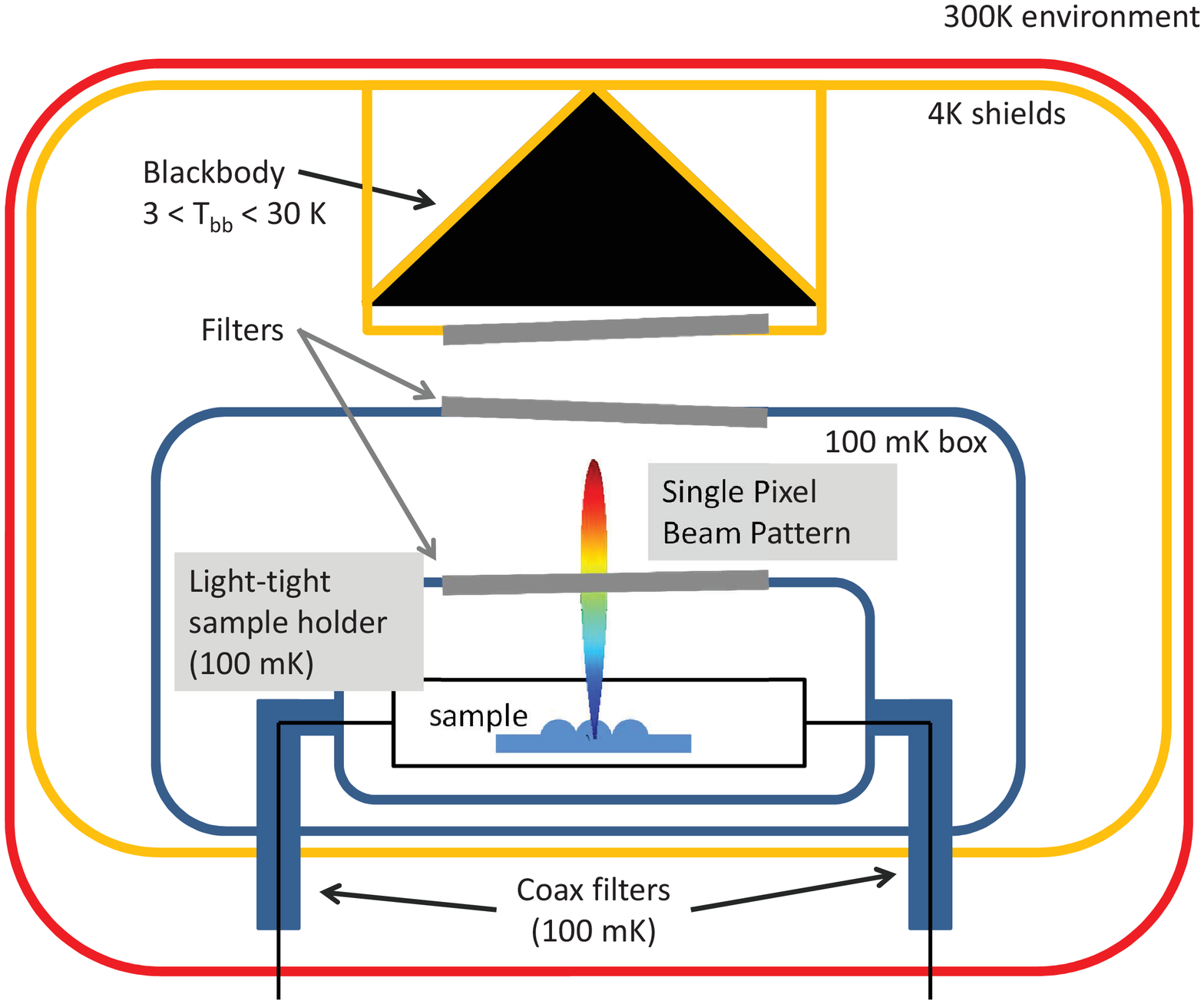} & \includegraphics[width = 0.25\textwidth]{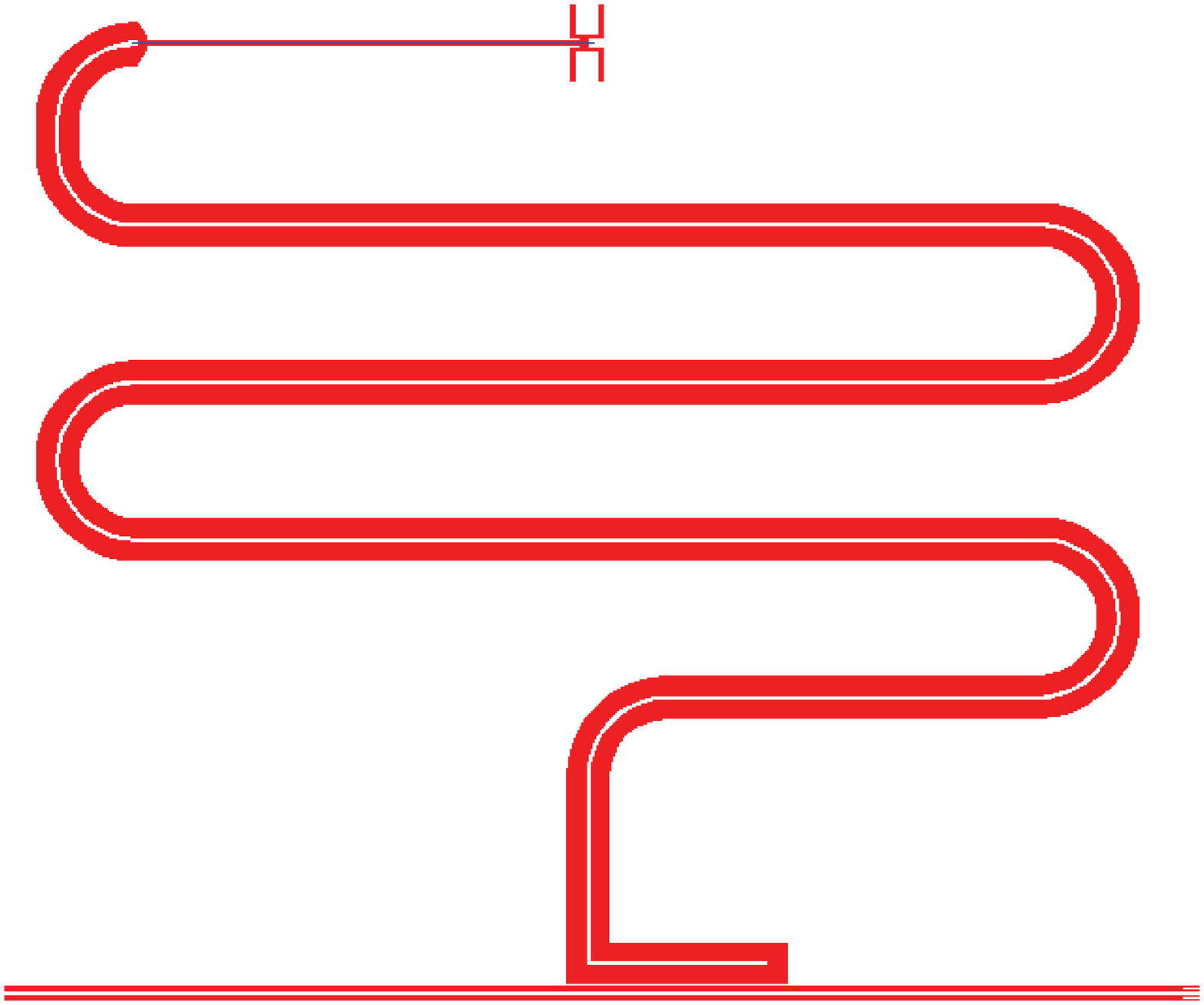} & \includegraphics[width = 0.25\textwidth]{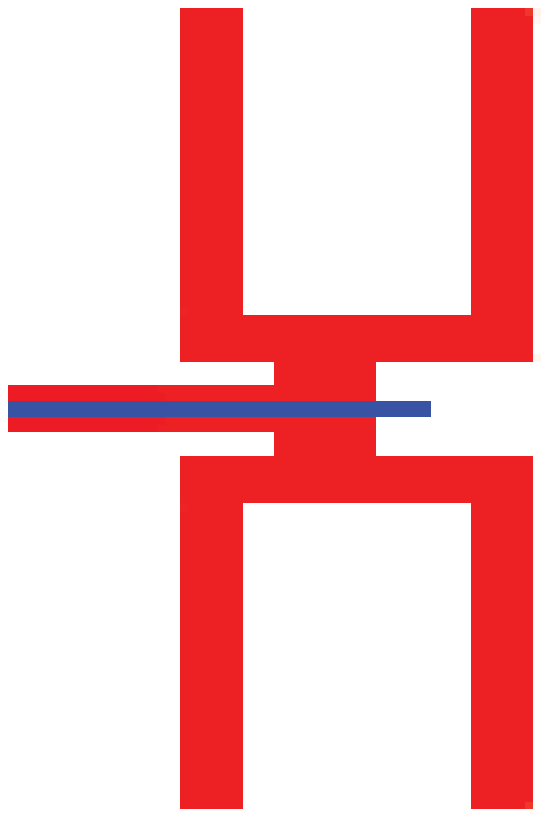}
   \end{tabular}
   \end{center}
   \caption[] 
   { \label{Figure:Setup} Schematic representation of the experimental setup \textit{Left:} Cartoon version of the measurement setup\cite{Baselmans2012b,deVisser2014b}. The MKID array is mounted in a sample box, which is surrounded by a light tight box. This entire box-in-a-box construction is thermally anchored to the pulse tube pre-cooled Adiabatic Demagnetization Refrigerator (ADR) and cooled to 100 mK. The box-in-a-box design in combination with the coax filters reduce the stray radiation from the 4K stage to $<60$ aW. A cryogenic blackbody source is placed on the 4 K stage created by the pulse tube and used to illuminate MKIDs. The power received by the MKIDs is restricted by a 20 mm aperture located 15 mm above the sample and 3 sets of metal mesh filters, which create a passband overlapping with the 850 GHz atmospheric window. Through the temperature control of the blackbody a variable illumination is achieved of $ 0.1 \ \mathrm{fW} < P_{\mathrm{850GHz}} <  2 $ pW. \textit{Middle:} Design of the second generation antenna-coupled hybrid NbTiN-Al MKIDs. A wide NbTiN (white) CPW resonator is used to minimize the two-level system noise contribution. At the shorted end, where the planar antenna is located, one millimeter of CPW is reduced in width and the central line is made from 40 nm Al (blue). A 4 by 4 array of hybrid MKIDs was made on a C-plane sapphire substrate (red).\textit{Right:} A zoom of the 850 GHz twin-slot antenna. Lenses on the other side of the sapphire substrate (red) focus the light on this antenna. This zoom also clearly shows the narrow Al central line (blue) connecting to the NbTiN (white).}
   \end{figure} 
The design of the second generation antenna coupled hybrid MKIDs is shown in the middle panel of Fig. \ref{Figure:Setup}. The device is a $L\approx 4.5$ mm long quarter wavelength CPW resonator consisting of two sections. The first section ($\sim 4$ mm), at the open end of the resonator, is a wide CPW made entirely from $\sim 500$ nm thick NbTiN. NbTiN has 10 dB lower TLS noise than conventional superconductors such as Al\cite{Barends2009}. The TLS noise is further reduced by the increased width of the CPW\cite{Barends2009}, $30 \ \mum$ and $10 \ \mum$ for the CPW gap and central line, respectively.\\
The second section ($\sim 0.5$ mm), at the shorted end of the resonator, is a narrow CPW with NbTiN groundplanes and a 40 nm thick Al central line. The Al is galvanically connected to the NbTiN central line and the NbTiN groundplane at the resonator short (Fig. \ref{Figure:Setup} right panel). The NbTiN is lossless for frequencies up to the gap $2\Delta_0/h = 1.1$ THz ($T_c \approx 14$ K). Any radiation with a frequency $0.09 < \nu < 1.1$ THz is therefore absorbed in the Al ($T_c = 1.28$ K) central line of the second section. The optically excited quasiparticles are trapped in the Al, because it is surrounded by a high gap superconductor. This quantum well structure confines the quasiparticles in the most responsive part of the MKID and allows us to maximize the response by minimizing the active volume. Therefore, we use a narrow CPW in section two, $2 \ \mum$ for both the CPW gap and central line. Using a narrow Al line at the shorted end of the MKID does not increase the TLS noise significantly, because of the negligible electric field strength in this part of the detector.\\
At the shorted end of the resonator light is coupled into the device through a single polarization twin-slot antenna. A zoom of the antenna is shown in the right-hand side panel of Fig. \ref{Figure:Setup}. The advantage of using antenna coupling is that it can be designed independently from the distributed CPW resonator. This gives antenna-coupled hybrid MKIDs the flexibility to be employed within the frequency range $0.09 < \nu < 1.1$ THz, where the NbTiN is lossless and the Al absorbs radiation. The disadvantage is that the antenna occupies $\sim 0.1\%$ of the total pixel footprint. To achieve a high filling fraction we use elliptical lenses to focus the light on the antennas. The second generation  hybrid MKIDs in these proceedings are coupled to a broad-band twin-slot antenna designed for observations in the $\nu=850$ GHz atmospheric window.\\
The hybrid MKID discussed in detail in these proceedings is part of a 4 by 4 array of hybrid MKID pixels with varying designs. The array was fabricated\cite{Lankwarden2012} on a C-plane sapphire substrate. All pixels are capacitively coupled to a single feedline. An array of laser machined Si lenses with a diameter of 2 mm is mounted on the array.
\section{Performance as Sub-millimeter Detector}
\subsection{Expermintal Setup}
The MKID array is evaluated using a pulse tube pre-cooled adiabatic demagnetization refrigerator with a box-in-a-box cold stage design\cite{Baselmans2012b} as shown in the left-hand side panel of Fig. \ref{Figure:Setup}. The two light tight boxes in combination with the feed through coax filters prevent 4 K radiation from entering the sample stage either directly or through the coax lines. In addition, the inside of the outer box is coated with carbon loaded epoxy and SiC grains to effectively prevent reflections of light entering through the filters. As a result, the array is fully enclosed in a 100 mK environment with the exception of a 20 mm aperture, which is located 15 mm above the MKID array. This aperture is isotropically illuminated by a large temperature controlled blackbody mounted on the 4K stage\cite{deVisser2014b}. Three stacks of metal mesh filters provide a minimum rejection of 40 dB at all wavelengths outside the 150 GHz bandpass centered on $\nu=850$ GHz. This setup allows us to create a variable loading on our MKIDs. We calculate the loading on the MKIDs, $P_{\mathrm{calc}}$, using\cite{Janssen2013}
\begin{equation}
P_{calc} = \frac{c^2}{4\pi} \int_{\nu} \int_{\Omega\in A_{ap}} \frac{F_{\nu}B_{\nu}(T_{BB})}{2\nu^2} \ C_{\nu}G_{\nu}(\Omega) \ d\Omega d\nu
\label{Eq:Pest}
\end{equation}
Here $c$ is the speed of light, $F_{\nu}$ the filter transmission, $C_{\nu}$ the coupling efficiency, and $B_{\nu}(T_{BB})$ Planck's law for a blackbody temperature $T_{BB}$. The factor $1/2$ takes into account that we receive only a single polarization. The second integral evaluates the gain pattern of the lens-antenna system, $G_{\nu}(\Omega)$ over the solid angles $\Omega$ that span aperture area, $A_{ap}$. The rest of the detector enclosure is a 100 mK absorber that has a negligible emission at 850 GHz. The coupling efficiency , $C_{\nu}$, which describes the reflection losses due to mismatches between the antenna and the resonator CPW, and the far field gain pattern, $G_{\nu}(\Omega)$, are obtained from a simulation of the complete lens-antenna system using CST Microwave Studio. The gain pattern of our lens-antenna system is included (to scale) in the lefthand side panel of Fig. \ref{Figure:Setup}. Due to the directivity of the beam and the large 20 mm aperture, the throughput of the setup is $0.81\lambda^2$. The main loss with respect to the $\lambda^2$ throughput of single moded receiver is the backlobe.\\
\subsection{Photon Noise Limited Performance}
   \begin{figure}[t]
   \begin{center}
   \begin{tabular}{c c}
   \includegraphics[width = 0.48\textwidth]{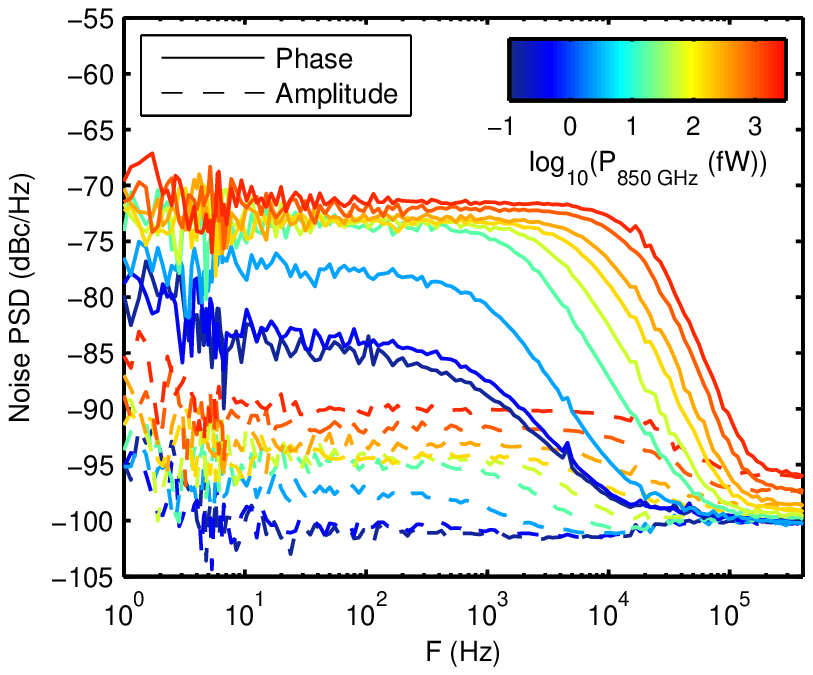} & \includegraphics[width = 0.48\textwidth]{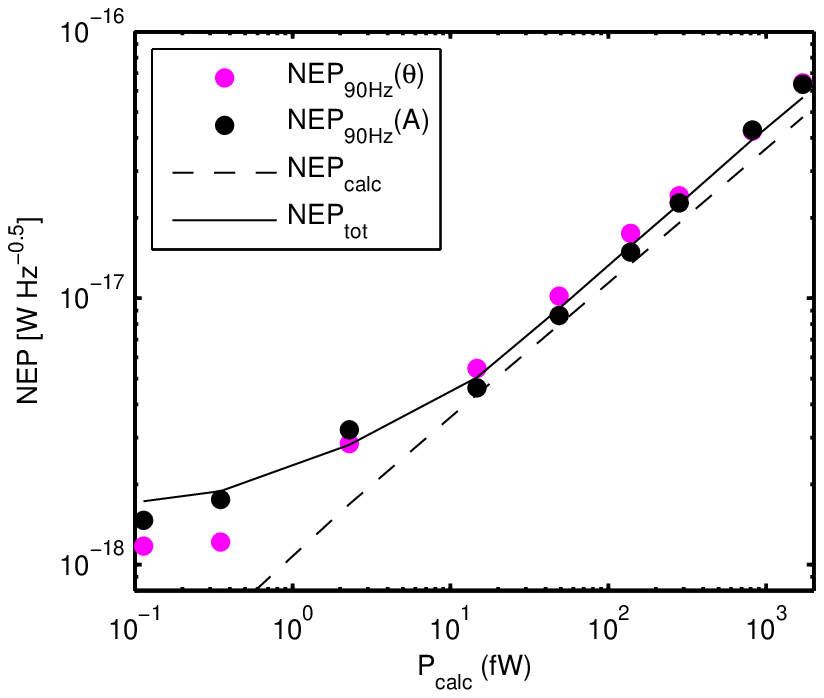}
   \end{tabular}
   \end{center}
   \caption[] 
   { \label{Figure:Spectra} \textit{Left:} The power spectral density of the amplitude (dashed) and phase (solid) noise measured under various optical loading. A white noise spectrum is observed for $P_{\mathrm{850GHz}} > 10$ fW. The roll-off above 1 kHz is due to the quasiparticle recombination time, which is reduced by an increasing optical load. Note that the photon noise level is 16 dB higher in phase readout. Nevertheless, at a given loading the NEP is the same for phase and amplitude readout as shown in the right-hand side panel. \textit{Right:} The optical NEP measured at a modulation frequency $F=90$ Hz, $NEP_{\mathrm{opt}}$, as a function of the estimated optical loading, $P_{\mathrm{calc}}$. For the loading levels where we are photon noise limited, $P_{\mathrm{calc}} > 10$ fW, the NEP in phase (magenta dots) and amplitude (black dots) readout is within a factor of 1.4, after correction for the amplifier noise contribution. The measured NEP follows the same slope as $NEP_{\mathrm{calc}} \propto \sqrt{P_{\mathrm{calc}}}$ (dashed black line). By fitting the relation between measured amplitude NEP and $NEP_{\mathrm{calc}}$ (solid black line) $\epsilon = 1.04 \pm 0.24$ is determined.}
   \end{figure} 
The left-hand side panel of Fig. \ref{Figure:Spectra} shows the amplitude and phase noise spectra measured for a typical device as a function of the optical power absorbed in the Al, $P_{\mathrm{850GHz}}$. Following the definition given by Janssen et al.\cite{Janssen2013} we see from these spectra that these devices are photon noise limited for $P_{\mathrm{850GHz}} > 10$ fW. In this regime the observed spectra are white and roll-off due to the quasiparticle recombination time. At loading levels below 0.5 fW the spectra are clearly dominated by the detector noise. In phase readout the spectra show the $1/\sqrt{F}$ dependence expected from TLS noise. The amplitude noise spectrum becomes white over the entire frequency range. The absence of a roll-off indicates we are limited by the cryogenic amplifier.\\
The right-hand side panel of Fig. \ref{Figure:Spectra} shows the optical Noise Equivalent Power (NEP) measured\cite{deVisser2014b} at a modulation frequency $F=90$ Hz, $NEP_{\mathrm{90Hz}}$, as a function of the estimated optical loading, $P_{\mathrm{calc}}$. For the loading levels where we are photon noise limited, $P_{\mathrm{calc}} > 10$ fW, the NEP in phase (magenta dots) and amplitude (black dots) readout are within a factor 1.4, after correction for the amplifier noise contribution. The measured NEP also follows a $NEP \propto \sqrt{P_{\mathrm{850GHz}}}$ relation. Both are expected for MKIDs\cite{Yates2011,Baselmans2012a} that are photon noise limited in both phase and amplitude readout.\\
Since the photon noise limited NEP of a detector only depends on the absorbed sub-mm radiation, we can verify $P_{\mathrm{calc}}$ by comparing the measured and expected NEP\cite{Janssen2013}. Fig. \ref{Figure:Spectra} (right panel) shows the expected photon noise limited NEP (dashed line), $NEP_{\mathrm{calc}}$,
\begin{equation}
NEP_{calc}(P_{\mathrm{calc}}) = \sqrt{2P_{calc}(h\nu(1+mB)+\Delta/\eta_{pb})}
\end{equation}
Here $h\nu$ is the photon energy of the incoming radiation, $(1+mB)$ the correction to Poisson statistics due to wave bunching, $\Delta$ the superconducting energy gap of the absorbing material and $\eta_{pb}= 0.4$ the pair breaking efficiency\cite{Guruswamy2014} for our 40 nm Al film. The measured NEP can then be described by
\begin{equation}
NEP_{\mathrm{tot}}^2 = NEP_{\mathrm{calc}}^2/\epsilon + NEP_{det}^2
\label{Eq:Epsilon}
\end{equation}
$NEP_{det}$ is a correction for the amplifier noise contribution, which we estimation from the NEP value at a modulation frequency of 300 kHz. We expect $\epsilon=1$, if the description of the optical power flow by Eq. \ref{Eq:Pest} is complete. As best fit value for $\epsilon$ we find $\epsilon = 1.04\pm0.24$.\\
This is a significantly larger uncertainty than found by Janssen et al.\cite{Janssen2013}. It is our hypothesis that this is due to the large 20 mm aperture in combination with various reflective surfaces (including the filter), which allows additional in-band radiation to reach the Al through multiple reflections. This is corroborated by the fact that $\epsilon$ increases for increasing $P_{\mathrm{calc}}$.\\
\subsection{Aperture Efficiency}
In the photon noise limited regime it is favorable to have a high optical or aperture efficiency\cite{Rohlfs2004}, $\eta_{A}$, because the observation time required to achieve a given signal-to-noise follows $t_{\sigma} \propto \eta_A^{-1}$. From the simulated far field beam pattern we can determine the aperture efficiency, $\eta_A$, which is mathematically defined as
\begin{equation}
\eta_A = \frac{A_e}{A} = \frac{\lambda^2G_{\nu}(\Omega_0)C_{\nu}}{4 \pi A}
\end{equation}
Here $A$ is the physical area covered by the pixel. $\lambda$ and $\nu$ are the wavelength and frequency of the observed radiation, respectively. $\Omega_{0}$ is the direction of the maximum gain. Using the circular area of the lenses $A=\pi$ $\mathrm{mm^2}$, $C_{350GHz}=0.91$ and the gain of the CST beam pattern at broadside, $G_{\nu}(\Omega_0) = 12.6$ dB, an aperture efficiency of 66\% is determined for a single pixel. The maximum achievable aperture efficiency of a circular antenna illuminated by a single moded gaussian beam is\cite{QuasiOptics} $\eta_A=0.80$. For the measured array the filling fraction of the square packing means we have a total array aperture efficiency of 52\%. Using an array with hexagonal packing the total array aperture efficiency can be increased to 60\%.\\
\subsection{Readout}
   \begin{figure}[t]
   \begin{center}
   \begin{tabular}{c}
   \includegraphics[width = 1.0\textwidth]{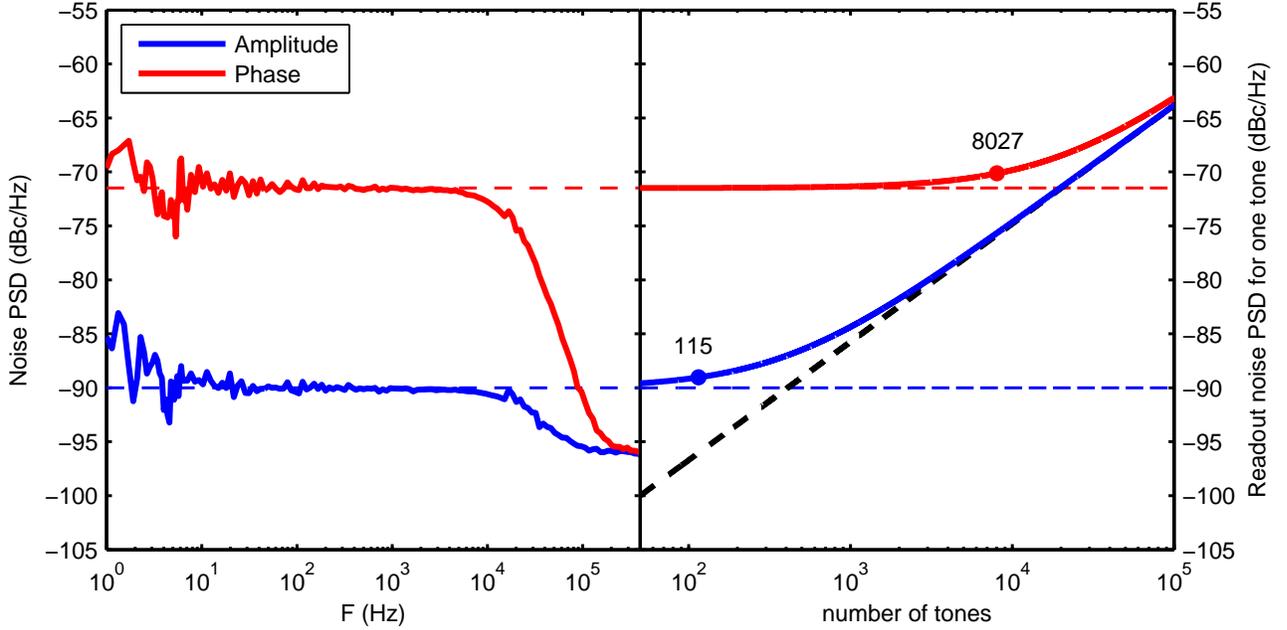}
   \end{tabular}
   \end{center}
   \caption[] 
   { \label{Figure:Readout} \textit{Left:} The measured amplitude (blue) and phase (red) noise spectra for an optical loading of 1 pW. For modulation frequencies $1<F<10^4$ Hz the spectra are white at a -71.4 dBc/Hz and -90 dBc/Hz level for phase and amplitude, respectively. \textit{Right:} The noise contribution (black dashed line) by a state-of-the-art readout system based upon the E2V EV10AQ190 ADC for a given number of readout tones. The solid lines indicate the sum of the photon and readout noise for phase (red) and amplitude (blue). If we allow for a 10\% degradation of observing time due to the noise added by the ADC alone, we can simultaneously operate 115 or 8027 tones in amplitude and phase readout, respectively.}
   \end{figure} 
Once photon noise limited performance is achieved the sensitivity in both phase and amplitude readout is equal. However, as shown in Fig. \ref{Figure:Spectra}, the photon noise level in phase readout is significantly higher than that in amplitude readout for the same NEP. Fig. \ref{Figure:Readout} shows the measured amplitude (blue) and phase (red) noise spectra for an optical loading of 1 pW. The photon noise level in phase readout is 18 dB higher than that in amplitude readout, thereby relaxing the dynamic range requirements of the readout electronics.  The right-hand side panel of Fig. \ref{Figure:Readout} shows the expected noise contribution (black dashed line) by a state-of-the-art readout system based upon the E2V EV10AQ190 ADC for a given number of readout tones. We used an effective number of bits (ENOB) of 8 for this calculation and a ADC bandwidth of 1.25 GHz. The calculation results shown in Fig. \ref{Figure:Readout} was verified experimentally. If we allow for a 10\% degradation of observing time or $\sim 1$ dB of added noise by the ADC alone, we can simultaneously operate 115 or 8027 tones in amplitude and phase readout, respectively. Because the number of usable tones is directly proportional to the number of pixels read simultaneously by the system, phase readout is favorable for the operation of large arrays.\\
It is also worth noting that the spectra are white for modulation frequencies $1<F<10^4$ Hz. For astronomical instruments readout (modulation) frequencies $F<10$ Hz are relevant.
\section{Discussion}
In these proceedings we have presented the performance of the second generation hybrid NbTiN-Al MKIDs. Compared to the hybrids presented by Janssen et al.\cite{Janssen2013} we have achieved a number of performance improvements, which make application in astronomically relevant instruments more plausible.
\begin{itemize}
\item We reduced the Al volume to increase the MKID response. Hereby we achieve a 5 dB higher photon noise level allowing 3 times more pixels.
\item The wider NbTiN CPW has further reduced the TLS noise. Initial comparison to literature\cite{Barends2009} suggests the measured TLS noise is contributed by the Al section.
\item Using a sapphire substrate increases the TLS noise by $\sim 7$ dB\cite{Barends2009}, but it improves the reproducibility of the fabrication process and enables a room temperature check of the through line integrity using a DC measurement of the resistance. This is an important help in the quality control during the fabrication of very large arrays, which can have a through line of $\sim 0.5$ m. 
\end{itemize}
In addition, the use of a 850 GHz shows the flexibility of the lens-antenna radiation coupling. The next step in this process will be coupling to on-chip filterbanks for sub-mm spectrometers such as DESHIMA\cite{Endo2012}. Compared to the hybrid MKIDs operating at 350 GHz the aperture efficiency did decrease by 10 percentage points. This is due to the absence of an AR coating on the lens, in combination with the large lens size.\\
\section{Conclusion}
In conclusion, we present the second generation hybrid NbTiN-Al MKIDs, which are photon noise limited in both phase and amplitude readout for loading levels $P_{\mathrm{850GHz}} \geq 10$ fW. The photon noise limited performance in phase allows us to simultaneously read out approximately 8000 pixels using state-of-the-art electronics. In addition, the use of a Si lens and a planar antenna gives these resonators the flexibility to operate within the frequency range $0.09 < \nu < 1.1$ THz. Given these specifications, hybrid NbTiN-Al MKIDs will enable astronomically usable kilopixel arrays for sub-mm imaging and moderate resolution spectroscopy.

\acknowledgments 
 
The authors thank D.J. Thoen and V. Murugesan for sample fabrication. T.M. Klapwijk and R.M.J. Janssen are grateful for support from NOVA, the Netherlands Research School for Astronomy, to enable this project. A. Endo is grateful for the financial support by NWO (Veni grant 639.041.023) and the JSPS Fellowship for Research Abroad. The work was in part supported by ERC starting grant ERC-2009-StG Grant 240602 TFPA (A.M. Baryshev). T.M. Klapwijk acknowledges financial support from the Ministry of Science and Education of Russia under contract No. 14.B25.31.0007 and ERC Advanced grant: No.339306 (METIQUM). 


\bibliographystyle{spiebib}   

\begin{thebibliography}{18}

\bibitem{Brown2004}
R.~L. {Brown}, W.~{Wild}, and C.~{Cunningham}, ``{ALMA - the Atacama large
  millimeter array},'' {\em Advances in Space Research}~{\bf 34}, pp.~555--559,
  Jan. 2004.

\bibitem{Scott2010}
D.~{Scott}, P.~{Barmby}, P.~{Bastien}, J.~{Cami}, E.~{Chapin}, J.~{Di
  Francesco}, M.~{Fich}, M.~{Halpern}, M.~{Houde}, G.~{Joncas}, D.~{Johnstone},
  P.~{Martin}, G.~{Marsden}, B.~{Matthews}, D.~{Naylor}, C.~{Barth
  Netterfield}, E.~{Peeters}, R.~{Plume}, A.~{Pope}, G.~{Schieven}, T.~{Webb},
  and C.~{Wilson}, ``{The Submillimetre Universe},'' {\em ArXiv e-prints} ,
  Aug. 2010.

\bibitem{Gusten2006}
R.~{G{\"u}sten}, L.~{\AA}. {Nyman}, P.~{Schilke}, K.~{Menten}, C.~{Cesarsky},
  and R.~{Booth}, ``{The Atacama Pathfinder EXperiment (APEX) - a new
  submillimeter facility for southern skies -},'' {\em Astronomy \&
  Astrophysics}~{\bf 454}, pp.~L13--L16, Aug. 2006.

\bibitem{Baars1987}
J.~W.~M. {Baars}, B.~G. {Hooghoudt}, P.~G. {Mezger}, and M.~J. {de Jonge},
  ``{The IRAM 30-m millimeter radio telescope on Pico Veleta, Spain},'' {\em
  Astronomy \& Astrophysics}~{\bf 175}, pp.~319--326, Mar. 1987.

\bibitem{Woody2012}
D.~{Woody}, S.~{Padin}, E.~{Chauvin}, B.~{Clavel}, G.~{Cortes}, A.~{Kissil},
  J.~{Lou}, P.~{Rasmussen}, D.~{Redding}, and J.~{Zolwoker}, ``{The CCAT 25m
  diameter submillimeter-wave telescope},'' in {\em Society of
  Photo-Optical Instrumentation Engineers (SPIE) Conference Series} {\bf 8444},
  Sept. 2012.

\bibitem{Day2003}
P.~K. {Day}, H.~G. {LeDuc}, B.~A. {Mazin}, A.~{Vayonakis}, and J.~{Zmuidzinas},
  ``{A broadband superconducting detector suitable for use in large arrays},''
  {\em Nature}~{\bf 425}, pp.~817--821, Oct. 2003.

\bibitem{Janssen2013}
R.~M.~J. {Janssen}, J.~J.~A. {Baselmans}, A.~{Endo}, L.~{Ferrari}, S.~J.~C.
  {Yates}, A.~M. {Baryshev}, and T.~M. {Klapwijk}, ``{High optical efficiency
  and photon noise limited sensitivity of microwave kinetic inductance
  detectors using phase readout},'' {\em Applied Physics Letters}~{\bf 103},
  p.~073505, 2013.

\bibitem{Gao2007}
J.~{Gao}, J.~{Zmuidzinas}, B.~A. {Mazin}, H.~G. {Leduc}, and P.~K. {Day},
  ``{Noise properties of superconducting coplanar waveguide microwave
  resonators},'' {\em Applied Physics Letters}~{\bf 90}, p.~102507, Mar. 2007.

\bibitem{Baselmans2012b}
J.~J.~A. {Baselmans}, S.~J.~C. {Yates}, P.~{Diener}, and P.~J. {de Visser},
  ``{Ultra Low Background Cryogenic Test Facility for Far-Infrared Radiation
  Detectors},'' {\em Journal of Low Temperature Physics}~{\bf 167},
  pp.~360--366, May 2012.

\bibitem{deVisser2014b}
P.~J. {de Visser}, J.~J.~A. {Baselmans}, J.~{Bueno}, N.~{Llombart}, and T.~M.
  {Klapwijk}, ``{Fluctuations in the electron system of a superconductor
  exposed to a photon flux},'' {\em Nature Communications}~{\bf 5}, p.~3130,
  Feb. 2014.

\bibitem{Barends2009}
R.~{Barends}, H.~L. {Hortensius}, T.~{Zijlstra}, J.~J.~A. {Baselmans}, S.~J.~C.
  {Yates}, J.~R. {Gao}, and T.~M. {Klapwijk}, ``{Noise in NbTiN, Al, and Ta
  Superconducting Resonators on Silicon and Sapphire Substrates},'' {\em IEEE
  Transactions on Applied Superconductivity}~{\bf 19}, pp.~936--939, June 2009.

\bibitem{Lankwarden2012}
Y.~J.~Y. {Lankwarden}, A.~{Endo}, J.~J.~A. {Baselmans}, and M.~P. {Bruijn},
  ``{Development of NbTiN-Al Direct Antenna Coupled Kinetic Inductance
  Detectors},'' {\em Journal of Low Temperature Physics}~{\bf 167},
  pp.~367--372, May 2012.

\bibitem{Yates2011}
S.~J.~C. {Yates}, J.~J.~A. {Baselmans}, A.~{Endo}, R.~M.~J. {Janssen},
  L.~{Ferrari}, P.~{Diener}, and A.~{Baryshev}, ``{Photon noise limited
  radiation detection with lens-antenna coupled microwave kinetic inductance
  detectors},'' {\em Applied Physics Letters}~{\bf 99}, p.~073505, 2011.

\bibitem{Baselmans2012a}
J.~J.~A. {Baselmans}, ``{Kinetic Inductance Detectors},'' {\em Journal of Low
  Temperature Physics}~{\bf 167}, pp.~292--304, May 2012.

\bibitem{Guruswamy2014}
T.~{Guruswamy}, D.~J. {Goldie}, and S.~{Withington}, ``{Quasiparticle
  generation efficiency in superconducting thin films},'' {\em Superconductor
  Science Technology}~{\bf 27}, p.~055012, May 2014.

\bibitem{Rohlfs2004}
K.~{Rohlfs} and T.~L. {Wilson}, {\em {Tools of radio astronomy}}, Springer,
  2004.

\bibitem{QuasiOptics}
P.~F. Goldsmith, {\em {Quasioptical systems : Gaussian beam quasioptical
  propagation and applications}}, Wiley - IEEE Press, 1998.

\bibitem{Endo2012}
A.~{Endo}, P.~{Werf}, R.~M.~J. {Janssen}, P.~J. {Visser}, T.~M. {Klapwijk},
  J.~J.~A. {Baselmans}, L.~{Ferrari}, A.~M. {Baryshev}, and S.~J.~C. {Yates},
  ``{Design of an Integrated Filterbank for DESHIMA: On-Chip Submillimeter
  Imaging Spectrograph Based on Superconducting Resonators},'' {\em Journal of
  Low Temperature Physics}~{\bf 167}, pp.~341--346, May 2012.

\end{thebibliography}

\end{document}